\documentclass[aps,prb,superscriptaddress,twocolumn,tightenlines]{revtex4-1}
\usepackage{amssymb}
\usepackage{graphicx}
\usepackage{natbib}
\usepackage{epstopdf}
\usepackage{dcolumn}% Align table columns on decimal point
\usepackage{bm}% bold math
\usepackage{makeidx}
\usepackage{enumerate}
\usepackage {graphicx}
\usepackage {amsmath}
\usepackage {amsfonts}
\usepackage {amssymb}
\usepackage {mathrsfs}
\usepackage {bbm}
\usepackage{subfigure}
\begin{document}
\title{Model of collective modes in three-band superconductors with  repulsive interband interactions}
\author{Valentin Stanev}
\affiliation{Materials Science Division, Argonne National Laboratory, Argonne, Illinois 60439}
\date{\today }

\begin{abstract}
I consider a simple model of a three-band superconductor with repulsive 
interband interactions. The frustration, associated with the odd number of bands, leads to the possible existence of an intrinsically 
complex time-reversal symmetry breaking (TRSB) order parameter. In such state the
fluctuations of the \emph{different} gaps are strongly coupled, and this leads to the development of novel collective modes, which mix phase and amplitude oscillations. I study these fluctuations using a simple microscopic model and derive the dispersion for two physically distinct modes, which are gapped by energy less than $2 \Delta$, and apparently present for all values of interband couplings.

%I consider a simple model of a three-band superconductor with repulsive interband interactions. In such a system the frustration associated with the odd number of gaps leads to the possible existence of intrinsically complex time-reversal symmetry breaking (TRSB) order parameter. I show that in this state the fluctuations of the \emph{different} gaps are strongly coupled, and this leads to the development of novel excitations, in which phase and amplitude oscillations are mixed. This is due to the non-trivial relative phase angle between the gaps.  The energy of these excitations is less than $2 \Delta$, and thus they are true collective modes of the system.

\end{abstract}
\maketitle

\section{Introduction}
%{\it Introduction} - 
Study of the collective modes in superfluids and superconductors has a long history. In the case of single-band systems there are 
two well-known modes -- Bogoliubov-Anderson \cite{Bogol, Ander} and Schmid \cite{Schmid}, representing oscillations of the phase or the amplitude of the order parameter, respectively. In superfluids Bogoliubov-Anderson is a gapless Goldstone mode, but in superconductors it couples to the electromagnetic field and is gapped by
energy of the order of the plasma frequency. The Schmid mode is always gapped by at least $2\Delta$ (and thus is always damped by interaction with quasiparticles).
In superconductors yet another mode, which couples phase oscillations with charge currents, appears close to $T_c$ \cite{CG} (it is known as Carlson-Goldman).

In multiband and multicomponent superconductors the situation is even richer. This was first realized by Leggett, \cite{Leggett} who considered a two-band superconductor, in which the bands are coupled via a Josephson-like 
term. In such system, apart from the Bogoliubov-Anderson and Schmid modes, another collective mode, representing oscillations of the relative phase of the two gaps, is possible (observation of this mode in  MgB$_2$ has been reported \cite{MgB2}). The Leggett mode is gapped in both superconductors and superfluids (for discussion of Leggett and Carson-Goldman modes in a dirty two-band superconductor see Ref. \onlinecite{Efetov}). 
There are analogs of this mode in systems with two-component order parameters (see, for example, Refs. 
\onlinecite{Ichioka, Balatski}), and it has been argued that there is a related mode in
strongly correlated superconductors \cite{Fionna}.
This mode exists if the different bands
are weakly coupled \emph{and} superconducting even in the case in which
the interband coupling vanishes. 
%Otherwise the oscillations in the relative phase are pair-breaking and thus overdamped.
In such case there are
two non-trivial mean-field (MF) superconducting states (with phase difference between the two gaps equal to $0$ or $\pi$). One of these states is metastable (which one depends on the sign of the interband coupling). When the interband
interactions become dominant
% ($\lambda_{12}>\lambda_{11}\lambda_{22}$, where $\lambda_{11}, \lambda_{22}$ are the intraband pairing interactions)
the metastable state and the Leggett mode disappear. \cite{Leggett}     
\begin{figure}[h]
\begin{center}$
\begin{array}{cc}
\includegraphics[width=0.35\textwidth]{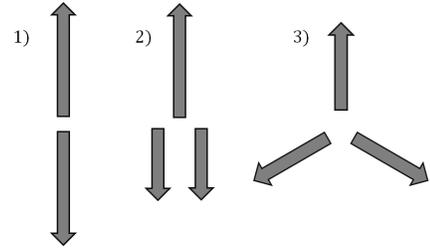}
\end{array}$
\end{center}
\caption{The possible superconducting states in a three-band system. The angles between the arrows represent the relative phase between the different gaps. On the left is the effective two-gap $s_{\pm}$ state, in the middle the three-gap $s_{\pm}$ state, and on the right the TRSB three-gap state (see Ref. \onlinecite{VS} for details). The complex state is possible if one or three of the interband pairing terms are repulsive.}
\label{Fig1}
\end{figure}

Recently, multiband superconductivity has attracted a lot of attention in connection with the iron-based high-temperature superconductors, 
since these materials may have up to four or five superconducting gaps on disconnected pieces of the Fermi surface \cite{FeAs}.
One interesting feature of these compounds
is the likely existence of strong \emph {repulsive} interband interactions.
% that superconductivity very likely originates from strong \emph {repulsive} interband interactions and is intrinsically a multiband effect. 
In the case of two-band superconductors such interactions lead to an unconventional state
with a relative minus sign between the gaps (the so-called $s_{\pm}$ state) \cite{Kondo}.
The three-band case is even more interesting, since in such system there is an intrinsic frustration, and several possible superconducting states compete \cite{Ng, VS, Ota, Tanaka, Dias, Hu1, Hu2, Babaev, sd} (schematically shown in Fig.\ref{Fig1}). First, the system
can effectively behave as a two-gap $s_{\pm}$ superconductor, with
 one of the bands remaining normal. There is also a three-gap $s_{\pm}$ state, with a $\pi$ phase difference between two of the gaps and the third one.
The most interesting possibility appears if the three interband interactions are roughly comparable in strength (and thus
 the two- and three-gap $s_{\pm}$ states are close in energy) -- then the system can develop an intrinsically complex order parameter, with non-trivial relative phases between the different gaps. Such a complex superconducting state breaks
the time-reversal symmetry.
Even if such order parameter is absent in bulk iron-based superconductors (which are more likely in some variation of the $s_{\pm}$ state \cite{Mazin}), a related order parameter can be induced by bringing in contact iron-based and conventional $s-$wave superconductors \cite{Ng, VS}. Frustration between the $s$-wave and the $s_{\pm}$ state at the interface can induce on
the pnictide side a relative phase, different from the bulk $\pi$ value.
Thus the TRSB state can be stabilized close to the boundary, at least under some circumstances (see also Ref. \onlinecite{Bobkov}).

\begin{figure}[h]
\begin{center}$
\begin{array}{cc}
\includegraphics[width=0.45\textwidth]{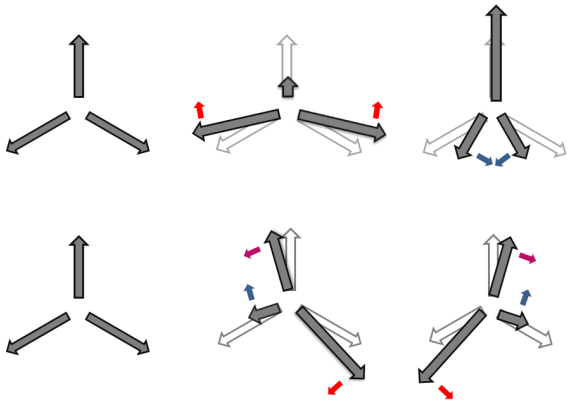}
\end{array}$
\end{center}
\caption{The two possible phase-amplitude modes around the TRSB state. The red/blue/purple arrows represent phase oscillations coupled with increase/decrease/no change of the amplitude of a given gap.}
\label{Fig2}
\end{figure}

In such intrinsically complex TRSB states there is a new phenomenon --  mixed phase-amplitude collective modes.  They  replace the more conventional Leggett modes and appear because phase and amplitude fluctuations of the gaps on the \emph{different} bands are coupled (unlike the case of multiband $s$ or $s_{\pm}$ states), due to the non-trivial relative phase angle. Recently these modes have been discussed in Refs. \onlinecite{Hu1, Babaev} using phenomenological 
Ginsburg-Landau (GL) theory \cite{GL}. Within this approach it was shown that such modes can have arbitrary low mass.  
In this paper I study the intergap coupling using alternative -- microscopic -- description. This approach is very similar in spirit to the one used in Refs. \onlinecite{Ota, Hu2}. There, however, the coupling between the phase and amplitude modes has been mostly neglected. In this work I show that such coupling is unavoidable and it creates true collective modes, which are gapped by energy less than $2 \Delta$, and seem to exist for all interband interactions. 

Note that coupling between amplitude and phase fluctuations is mandated by the Galilean invariance even in the single-gap case. However, these mixing terms can be 
neglected in the long-wavelength limit. \cite{Aitchison}
In contrast, the coupling between phase and amplitude fluctuations of the
 \emph {different} gaps in the TRSB state has a different physical
origin and survives even in the long-wavelength limit.   

\section{Pair susceptibility and linear response}      
%{\it Pair susceptibility and collective modes} - 
To study the problem I start from a simple generalization of the Bardeen-Cooper-Schrieffer (BCS) model for a three-band case:
\begin{eqnarray}
\mathcal{H}=\sum_{\bf {p}} \xi_{\bf {p}, \alpha} c^{\dagger}_{\bf {p} \alpha} c_{\bf {p} \alpha} + \sum_{\bf {p}\bf {p'}\bf {q}, \alpha \beta} V^{\alpha \beta}_{\bf {p}\bf {q}}  c^{\dagger}_{\bf {p} \alpha} c^{\dagger}_{\bf {p'} \alpha} c_{\bf {q} \beta} c_{\bf {q'} \beta},
\end{eqnarray}  
where $\alpha$ is a band index, the spin indexes are suppressed for the moment, and $ V^{\alpha \beta}_{\bf {p}q}$ is the pairing interaction matrix, which has both intraband ($\alpha=\beta$) and interband ($\alpha\neq\beta$) components. The TRSB state exists when one or three of the interband interactions are repulsive, and for concreteness I concentrate on the latter case. The intraband interactions can be both negative (attractive) or positive (repulsive), but even then the system can become superconducting, if the interband interactions are sufficiently strong.  

I decouple the interaction terms in a standard BCS fashion.
%, assuming square well potentials for $ V^{\alpha \beta}_{\bf {p}\bf {q}}$. 
This leads to multiband weak-coupling Hamiltonian
 \begin{eqnarray}
\mathcal{H}=\sum_{\bf {p}} \xi_{\bf {p} \alpha} c^{\dagger}_{\bf {p} \alpha} c_{\bf {p} \alpha} + \sum_{\bf {p}\bf {p'},\alpha} \Delta_{\bf {p},\alpha}  c^{\dagger}_{\bf {p+p'} \alpha} c^{\dagger}_{\bf {-p'} \alpha} +\mathrm{h.c.},
\end{eqnarray}  
where from the self-consistency condition
\begin{eqnarray}
\Delta_{\bf {p}, \alpha} = V^{\alpha \beta} \sum_{\bf {q}\beta} \langle c_{\bf {p+q}, \beta, \uparrow} c_{\bf {-q}, \beta, \downarrow}\rangle \nonumber
\end{eqnarray}
(restoring the spin indexes).
I will assume that the bands are identical -- a rather artificial case, but it simplifies the calculations considerably (for several examples of non-iron based compounds for which this model may apply see Ref. \onlinecite{Agterberg}). Thus $\xi_{p, 1}=\xi_{p, 2}= \xi_{p, 3}\equiv \xi$,  $V^{12}=V^{13}=V^{23}\equiv J$, $V^{11}=V^{22}=V^{33}\equiv V$. It is useful to expressed $\Delta$'s through the auxiliary variables $d$:
\begin{eqnarray}
d_{{\bf p},\alpha} = \sum_{{\bf q}}\langle c_{{\bf p+q}, \alpha, \uparrow} c_{{\bf -q}, \alpha, \downarrow}\rangle, \nonumber
\label{axvar}
\end{eqnarray}
for which it is straightforward to get:
\begin{eqnarray}
d_1=\frac{J(\Delta_2+\Delta_3)- (J+V)\Delta_1}{(J-V)(2J+V)}
\end{eqnarray} 
and analogous expressions for $d_2$ and $d_3$.
 
To study the fluctuations around the TRSB state I use standard linear response theory \cite{Kulik}. 
In the case of identical bands the free energy below $T_c$ is minimized by the 
(uniform) state $\Delta_{1,MF}=\Delta_0$, $\Delta_{2,MF}=\Delta_0 e^{2 \pi i/3}$, $\Delta_{3,MF}=\Delta_0 e^{4 \pi i/3}$ (up to an arbitrary overall phase)\cite{VS}. The other mean field solutions -- the two and three gap $s_{\pm}$ states -- are degenerate and separated from the TRSB state by finite energy.
I introduce small fluctuations around the TRSB state, and get 
\begin{eqnarray}
\Delta_{1}&=&\Delta_0+\rho_1 + i\theta_1, \nonumber\\
\Delta_{2}&=&\Delta_0 e^{2 \pi i/3}-\frac{1-i\sqrt{3}}{2}\rho_2 -\frac{\sqrt{3}+i}{2}\theta_2,\nonumber\\
\Delta_{3}&=&\Delta_0 e^{4 \pi i/3}-\frac{1+i\sqrt{3}}{2}\rho_3 +\frac{\sqrt{3}-i}{2}\theta_3,\nonumber
\end{eqnarray}

Here $\rho({\bf r}, t)$ and $\theta({\bf r}, t)$ represent the amplitude and phase fluctuations, respectively (note that both have dimension of energy).  With this the Hamiltonian can be split into two parts -- $\mathcal{H}=\mathcal{H}_0+\mathcal{H}^{\prime}$, where $\mathcal{H}_0$ contains all the one-particle and mean-field terms, and $\mathcal{H}^{\prime}$ can be regarded as a (time-dependent) perturbation: 
\begin{widetext}
\begin{eqnarray}
\mathcal{H}^{\prime}=\sum_{\bf{pq}}&& \Big [\Psi_{{\bf p+q},1}^{\dagger}(\rho_1 \sigma_1  +\theta_1 \sigma_2  + e \phi \sigma_3)\Psi_{{\bf p},1} + \Psi_{{\bf p+q},2}^{\dagger}\left(\frac{-\sigma_1+ \sqrt{3} \sigma_2}{2} \rho_2 -\frac{\sqrt{3} \sigma_1 + \sigma_2}{2} \theta_2 + e \phi \sigma_3\right)\Psi_{{\bf p},2} + \nonumber\\
&&\Psi_{{\bf p+q},3}^{\dagger}\left(\frac{\sigma_1 + \sqrt{3} \sigma_2}{2} \rho_3 + \frac{\sqrt{3} \sigma_1 - \sigma_2}{2} \theta_3 + e \phi \sigma_3\right)\Psi_{{\bf p},3}\Big ] e^{-i \omega t},
\end{eqnarray} 
\end{widetext}
where I used Nambu 2-spinors $\Psi^{\dagger}_{\bf{p}, \alpha}=[c^{\dagger}_{\bf{p}, \alpha, \uparrow},c_{\bf{p}, \alpha, \downarrow} ]$, the Pauli matrices $\sigma_i$, momentum-frequency representation of $\theta$ and $\rho$, and added scalar potential $\phi$. 

 The response of the superconducting state 
to a perturbation ${\bf h}_{\alpha}=[\rho_{\alpha}, \theta_{\alpha},  e \phi ]$ around its mean field value can be written as:
\begin{eqnarray}
g^{i}_{\alpha}({\bf p}, \omega)=\sum_j\Pi_{\alpha \alpha}^{ij}({\bf p}, \omega) h_{\alpha}^j({\bf p}, \omega),
\label{resfun}
\end{eqnarray}
where  
$g^{i}_{\alpha}=\mathrm{Tr}\sum{\langle\Psi^{\dagger}_{\bf{p}, \alpha} \sigma_i\Psi_{\bf{p+q}, \alpha} \rangle}  e^{-i \omega t} $.
Using Matsubara frequencies formalism
the polarization tensor is:  
\[ 
\Pi_{\alpha \alpha}^{ij}({\bf p}, \omega)=\mathrm{Tr}\  T \sum_{{\bf q, q'}, \nu}\sigma_i  \hat{\mathcal{G}}^{\alpha \alpha}_{{\bf q, q^{\prime}}}(\nu)\sigma_j \hat{\mathcal{G}}^{\alpha \alpha}_{{\bf q-p, q^{\prime}-p}}(\omega-\nu) 
\]
where $\hat{\mathcal{G}}$ is the electronic Green's functions in a superconductor:
\[
 \hat{\mathcal{G}}^{\alpha \alpha}_{{\bf p, p^{\prime}}}(\omega)=-\langle\Psi^{\dagger}_{{\bf p}, \alpha}
\Psi_{{\bf p^{\prime}}, \alpha} \rangle = -\frac{\xi_{p,\alpha} \sigma_3 + i \omega \sigma_0 - \Delta_{\alpha} \sigma_1}{\omega^2 + |\Delta_{\alpha}|^2 + \xi_{p,\alpha}^2} \delta_{{\bf p, p^{\prime}}}
\] 
In writing the linear response in Eq. \eqref {resfun} I have used 
the fact that $\hat{\mathcal{G}}$ is \emph {diagonal} in band indexes and there are no $\Pi_{\alpha \beta}^{ij}$  terms with $\alpha\neq\beta$. This means that the coupling between fluctuations in different bands enters only through the self-consistency equations. Furthermore, in the absence of supercurrents, the matrix elements coupling amplitude to phase fluctuations of the same gap, and to the scalar potential field vanish
in the long-wavelength limit ($\Pi^{12}_{\alpha \alpha}=\Pi^{21}_{\alpha \alpha}=\Pi^{13}_{\alpha \alpha}=\Pi^{31}_{\alpha \alpha}=0$ for each band). Finally, in the model with identical bands 
$\Pi_{11}^{ij}=\Pi_{22}^{ij}=\Pi_{33}^{ij}\equiv \Pi^{ij}$.

With this Eq. \eqref{resfun} can be written in a relatively simple form. Using the $d$ variables (note that Eq. \eqref{axvar} holds for each $({\bf p}, \omega)$ component) the equation for the purely real amplitude fluctuations 
of the first gap can be written as
\begin{eqnarray}
2 \mathrm{Re}(d_1({\bf p}, \omega))- \Pi^{11}\rho_1({\bf p}, \omega)=0. \nonumber
\end{eqnarray}
It is convenient to define 
\begin{equation}
A=\frac{J}{\nu_0(J-V)(2J+V)}, \ \ B=\frac{V}{\nu_0(J-V)(2J+V)}, \nonumber
\end{equation} 
and write the expression for $d_1$:
\begin{align}
 \mathrm{Re}(d_1)&=-\nu_0(A+B)\mathrm{Re}(\Delta_1) +
 \nu_0A(\mathrm{Re}(\Delta_2)+\mathrm{Re}(\Delta_3))\nonumber\\
&=-\nu_0(A+B)\rho_1
-\frac{\nu_0 A}{2} [(\rho_2+\rho_3)+\sqrt{3}(\theta_2-\theta_3)],\nonumber
\end{align} 
where $\nu_0$ denotes the density of states.

The equation for the phase fluctuations of $\Delta_1$ is
 \begin{eqnarray}
2 \mathrm{Im}(d_1({\bf p}, \omega))-\Pi^{22}({\bf p}, \omega)\theta_1({\bf p}, \omega) - \Pi^{23}({\bf p}, \omega) e\phi({\bf p}, \omega)=0, \nonumber
\end{eqnarray}
with:
\begin{align}
 \mathrm{Im}(d_1)&=-\nu_0(A+B)\theta_1
+\frac{ \nu_0 A}{2} [\sqrt{3}(\rho_2-\rho_3)-(\theta_2+\theta_3)].\nonumber
\end{align}
 I also define $Q^{23}\equiv \Pi^{23}/\nu_0$, $ \Pi^{ii}/\nu_0 + 2\int d\xi \tan(\sqrt{\xi^2+\Delta_0^2}/2T)/\sqrt{\xi^2+\Delta_0^2}\equiv Q^{ii}$, and express the integral through $A$ and $B$ using the gap equations (for details see Refs. \onlinecite{Kulik, Sharapov}). Utilizing these definitions and the equivalence of the bands Eq. \eqref{resfun} can be written in compact matrix form:
\begin{widetext}
\begin{eqnarray}
\begin{bmatrix}
 Q^{11}-2A  & A   & A     & 0    & \sqrt{3}A     &   -\sqrt{3} A   &0  \\
 A      & Q^{11}-2A & A     & -\sqrt{3} A     & 0 &   \sqrt{3} A   &0  \\
  A    &  A   &Q^{11}-2A  & \sqrt{3}A     & -\sqrt{3}A      &   0  & 0   \\
 0    &    -\sqrt{3} A   & \sqrt{3} A    & Q^{22}-2A &   A   & A  &   Q^{23}  \\
  \sqrt{3} A    &  0   &   -\sqrt{3} A  &  A  & Q^{22}-2A  &   A  &  Q^{23}  \\
 -\sqrt{3} A    &  \sqrt{3} A   &   0  &    A  &   A   & Q^{22}-2A & Q^{23}  \\
       0  &   0    &    0  & Q^{32} &Q^{32}&Q^{32}&  Q^{33}-p^2/(4 \pi e^2) \\
 \end{bmatrix} 
\begin{bmatrix}
       \rho_1 \\
       \rho_2\\
       \rho_3\\
       \theta_1 \\
       \theta_2 \\
       \theta_3 \\
       e \phi
   \end{bmatrix} =0
 \end{eqnarray}
\end{widetext}
%The normal modes of the system are given by the eigenvectors with real frequency. 
The Poisson equation has been used in the last element of the lowest matrix row. Note that the coupling constants $J$ and $V$ appear in the matrix only as a combination in $A$, which controls all intergap fluctuation couplings. This is due to the high symmetries of the model with identical bands we are considering. In the case of non-identical bands the couplings between the phase-phase and phase-amplitude fluctuations are generally different and the matrix becomes $19\times 19$.

\section{Collective modes}
The collective modes of the system can be obtained after analytically continuing to real frequencies. They are given by the eigenvectors of the matrix with eigenvalues equal to zero.
The explicit expressions are quite unwieldy, but analyzing them reveals a relatively simple picture. There are six eigenvectors, and two of them represent 
locked-in oscillations of the phase or amplitude of the three gaps, which are trivial generalization of the single-gap Bogoliubov-Anderson and Schmid 
modes. As in the single-gap case, the Bogoliubov-Anderson mode couples to the electromagnetic field (which gives it a mass), while the Schmid mode does not (but is nonetheless gapped by $2\Delta_0$). The remaining eigenvectors represent
the new coupled amplitude and phase fluctuations modes (see Fig. \ref{Fig2}). There are four   eigenvectors and two eigenvalues, so each eigenvalue is two-fold degenerate.
% -- thus there are two distinct modes with identical dispersions. 
The energy spectrum of the modes can be obtained by solving the equation 
%\begin{figure}[h]
%\includegraphics[width=0.45\textwidth]{CollectMode5.eps}
%\caption{$A$ as a function of $J$, with $V$ fixed at $-0.4$. Note that both positive and negative values are possible, and there is infinite discontinuity at $J=V/2$.The dashed line is set at $0.5$.}
%
%\end{figure}
\begin{equation}
- 6 A + Q^{11}+ Q^{22} \pm \sqrt{36 A^2 + (Q^{11} - Q^{22})^2} =0.
\label{Disp1}
\end{equation} 

For simplicity I consider only the case $T=0$ (no thermally excited quasiparticles). If $Q^{ii}$ are small they can be expanded to the lowest order in $p$ and $\omega$ \cite{Kulik, Sharapov}:
\begin{eqnarray}
Q^{11}\approx-\frac{\omega^2- c^2 p^2-4\Delta_0^2}{4\Delta_0^2}, \ \ 
Q^{22}\approx-\frac{\omega^2-c^2 p^2}{4\Delta_0^2},\ \
 Q^{23}\approx\omega,
\nonumber
\end{eqnarray}
where $c=v_F/\sqrt{3}$ gives the mode velocity. 
Using these results Eq.\eqref{Disp1} simplifies to 
\begin{eqnarray}
\omega^2= 2 \Delta_0^2\left(- 6 A+ 1 \pm \sqrt{36 A^2+1}\right)+c^2 p^2
\label{Disp2}
\end{eqnarray}
Obviously, the first term on the right side gives the energy gap $\omega_0$ of the mixed modes. 
\begin{figure}[h]
%\begin{center}$
%\begin{array}{cc}
\includegraphics[width=0.4\textwidth]{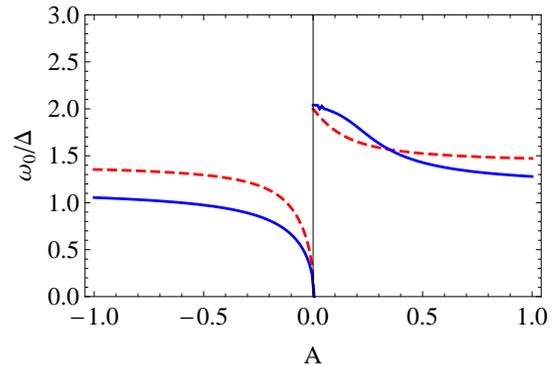}
%\includegraphics[width=0.23\textwidth]{CollectMode42.eps} &
%\includegraphics[width=2.5in]{CollectMode3.eps}
%\end{array}$
%\end{center}
\caption{The energy gap of the mixed modes as function of $A$, as derived from 
 Eqs.\eqref{Disp1} (blue curve) and  Eqs.\eqref{Disp2}  (red dashed curve). Apparently the modes can exist both for small and large $J$ (negative and positive $A$).}
\label{DispFig}
\end{figure}

Before considering the eigenmodes let us look at the possible values of $A$. As seen from its  definition, it can be positive or negative. When $V$ is finite and attractive (i.e. negative) $A$ has infinite discontinuity at $J=|V|/2$ and is negative/positive for smaller/larger values of $J$. For the case in which both interactions are repulsive (positive) superconductivity can exist only for $J>V$, and $A$ is always positive and decreasing function of $J$.      

To obtain $\omega_0$ these equations have to be solved. Equation \eqref{Disp1} can only be solved numerically, while  Eq.\eqref{Disp2} allows analytic treatment. In both cases it is easy to see that the mixed modes are well
defined ($\omega_0$ is real and below $2 \Delta_0$) for positive (negative) $A$ for the plus (minus) sign in
the equation. Thus for each $A$ there are two distinct collective modes (schematically shown on Fig. \ref{Fig2}) with identical dispersion.
The exact and the approximate solutions for $\omega_0$ are plotted on Fig. \ref{DispFig} and from it we see that these modes appear to exist for any given $A$ in the interval $(-\infty, \infty)$. As the coupling between the different gaps decreases ($J\rightarrow 0$, small negative $A$) $\omega_0$ also becomes smaller. 

Now let me compare these modes with the well-known collective modes in one- and two-band superconductors. Note that despite the fact that the mixed modes involve amplitude oscillations, they are gapped by less than $2 \Delta_0$. This is quite surprising and is entirely due to the phase-amplitude coupling. In contrast, purely amplitude modes cannot have energy less then twice the gap. The dispersion of the mixed modes is similar to that of the Leggett mode, but again there are important differences. As the results above indicate, the mixed modes are present for all values of $J$, in sharp contrast to the Leggett mode, which only exists for weak interband interactions. There is one crucial difference between the two- and the three-band superconductors, which may explain this  -- in the former the number of the mean-field solutions reduces from two to one when the interband interaction starts to dominate. In the three-band case  such reduction does not occur, and there can be three mean-field order parameters, even if the intraband interactions are set to zero. However, the reader should be warned that using the above results in the region of small and positive $A$ may be problematic, for at least two reasons. First, this region is formally outside of the regime of validity of the weak-coupling calculation (since in this region $J$ is large). The second and more physical reason is that when interband interactions dominate it is unclear whether well defined relative phase even exist.

Several features of the above results are due to the peculiarity of the particular case with identical bands. First, the electromagnetic potential does not couple to the mixed modes. In the general case such coupling should be expected, since phase fluctuations are conjugate to density fluctuations, and changes in the relative density produce charge imbalance, which couples to the field. Second, there are two degenerate in energy, but physically different modes (shown in Fig. \ref{Fig2}). The degeneracy is due to the fact the two metastable $s_{\pm}$ states (states 1 and 2 in Fig. \ref{Fig1}) have the same energy. In the case of non-identical bands these states split, and there are two different modes with distinct dispersions.   

More detailed and sophisticated calculations are necessary to study the possible existence of such modes in real superconductors, such as iron pnictides. These modes can be observed experimentally, for example by Raman scattering, \cite{MgB2, Hu2} and their presence can be used as a probe of the TRSB state.    

\section{Conclusions}
%{\it Conclusions}
In conclusion, I have considered a simple microscopic model of a three-band superconductor
with repulsive interband interactions. In such a system the frustration associated with the odd number of bands can lead to an intrinsically 
complex TRSB state. In this state two distinct collective modes develop, which necessarily mix the phase and amplitude oscillations of the \emph {different} gaps. These modes appear to be gapped by energy less than $2 \Delta$ in the cases of both weak and strong interband couplings.

\section*{ACKNOWLEDGMENTS}
I am very grateful to M. R. Norman, A. Levchenko, and A. E. Koshelev for useful discussions. This work was supported by the Center for
Emergent Superconductivity, a DOE Energy Frontier Research Center, Grant No. DE-AC0298CH1088. This manuscript was completed while enjoying the hospitality of the Aspen Center for Physics, supported by NSF under Grant No. 1066293.

 \bibliographystyle{apsrev}

\begin{thebibliography}{10}
\bibitem {GL} Note, however, that the applicability of the GL approach  
for the study of the TRSB state is not guaranteed. The reason is that this state can appear deep inside the superconducting region, where GL theory is not valid.
\end{thebibliography}

\end{document}